\documentclass[pra]{revtex4}
 \usepackage{amssymb} \usepackage{graphicx}

\begin{document}
 \title{Rigidity of Killing-Yano and conformal Killing-Yano superalgebras}

\author{\"Umit Ertem}
 \email{umitertemm@gmail.com}
\address{Department of Physics,
Ankara University, Faculty of Sciences, 06100, Tando\u gan-Ankara,
Turkey\\}

\begin{abstract}

Symmetry algebras of Killing vector fields and conformal Killing vectors fields can be extended to Killing-Yano and conformal Killing-Yano superalgebras in constant curvature manifolds. By defining $\mathbb{Z}$-gradations and filtrations of these superalgebras, we show that the second cohomology groups of them are trivial and they cannot be deformed to other Lie superalgebras. This shows the rigidity of Killing-Yano and conformal Killing-Yano superalgebras and reveals the fact that they correspond to geometric invariants of constant curvature manifolds. We discuss the structures and dimensions of these superalgebras on $AdS$ spacetimes as examples.

\end{abstract}

\maketitle

\section{Introduction}

Killing vector fields are generators of the symmetry algebra of a manifold which corresponds to the Lie algebra of isometries. They can be generalized to higher rank tensors called the hidden symmetries of the background manifold. The generalizations to symmetric tensors are called Killing tensors and they are related to the generalized constants of motion of the geodesic equation \cite{Benn}. The antisymmetric generalizations of Killing vector fields to higher degree differential forms are called Killing-Yano (KY) forms \cite{Yano}. They also have relations with the geodesic constants of motion \cite{Hughston Penrose Sommers Walker} and can be used in the separability of the Hamilton-Jacobi and Dirac equations in various backgrounds \cite{Cariglia Krtous Kubiznak, Cariglia}. The symmetry operators of the massive Dirac equation are constructed out of KY forms in curved backgrounds \cite{Benn Kress, Acik Ertem Onder Vercin1} and the odd degree KY forms also play the main role in the construction of the symmetry operators of geometric Killing spinors \cite{Ertem1} while the $p$-form Dirac currents of geometric Killing spinors correspond to KY forms or their Hodge duals \cite{Acik Ertem1, Acik}. Similarly, conformal Killing vector fields generate the conformal algebra of a manifold which corresponds to the Lie algebra of conformal isometries. They can also be generalized to higher rank symmetric and antisymmetric tensors and the antisymmetric generalizations to higher degree differential forms are called conformal Killing-Yano (CKY) forms. CKY forms are used in the construction of the symmetry operators of the massless Dirac equation \cite{Benn Charlton}. They also generate the symmetry operators of twistor and gauged twistor spinors \cite{Ertem2, Ertem3} and the $p$-form Dirac currents of twistor spinors correspond to CKY forms \cite{Acik Ertem1}.

Killing vector fields and geometric Killing spinors constitute a Lie superalgebra structure in different backgrounds which is called the Killing superalgebra \cite{Klinker, Alexeevsky Cortes Devchand Van Proeyen, OFarrill HackettJones Moutsopoulos Simon}. The even part of the superalgebra is the Lie algebra of Killing vector fields and the odd part is the set of geometric Killing spinors. In Minkowski spacetime, this superalgebra corresponds to the Poincar\'{e} superalgebra. In supergravity backgrounds which are the solutions of the field equations of the bosonic supergravity theories in different dimensions, supergravity Killing superalgebras can be constructed from Killing vector fields and supergravity Killing spinors \cite{OFarrill Meessen Philip, OFarrill HackettJones Moutsopoulos}. For the case of conformal Killing vector fields and twistor spinors, one can also construct Lie superalgebras which are called conformal superalgebras by adding some extra R-symmetry generators \cite{de Medeiros Hollands}. The structure of Killing and conformal superalgebras can be extended to include KY forms and CKY forms respectively by using the graded Lie algebra structure of hidden symmetries and the symmetry operators of Killing and twistor spinors \cite{Ertem1, Ertem2, Ertem5}. Although they constitute a superalgebra structure, they do not correspond to Lie superalgebras. On the other hand, KY forms and CKY forms satisfy some other Lie superalgebra structures of their own. The Lie bracket of Killing vector fields can be generalized to higher degree differential forms and it is called the Schouten-Nijenhuis (SN) bracket \cite{Schouten, Nijenhuis}. The SN bracket generates the Lie superalgebra of KY forms that is the KY superalgebra in constant curvature manifolds \cite{Kastor Ray Traschen, Acik Ertem2}. The even part consists of the Lie algebra of odd degree KY forms and the odd part of the superalgebra is the set of even degree KY forms. Similarly, the CKY superalgebra can be defined from a generalization of the SN bracket which is called the CKY bracket in constant curvature and Einstein manifolds \cite{Ertem4}. Again, the even part of the superalgebra consists of the odd degree CKY forms and the odd part is the set of even degree CKY forms.

A cohomology theory on a Lie superalgebra can be defined by using the bilinear operation of the Lie superalgebra \cite{Leites, Fuks}. By defining a filtration and $\mathbb{Z}$-gradation on a Lie superalgebra, one can obtain the generalized Spencer cohomology theory of Lie superalgebras \cite{Cheng Kac}. The importance of the Lie superalgebra cohomology stems from the fact that the possible deformations of a Lie superalgebra are related to its cohomology groups. If the second cohomology group is non-trivial, then the Lie superalgebra brackets can be deformed to obtain another Lie superalgebra. If it is trivial, then the Lie superalgebra cannot be deformed to another one and it is said that the Lie superalgebra is rigid. The Lie superalgebra deformations are important in supergravity theories and recently it is shown that the filtered deformations of the subsuperalgebras of the Poincare superalgebra can give rise to the supergravity Killing superalgebras in various dimensions \cite{OFarrill Santi1, OFarrill Santi2, de Medeiros OFarrill Santi, OFarrill Santi3}. Moreover, the field equations of bosonic supergravity theories can also be obtained from the calculations of the Spencer cohomology groups \cite{OFarrill Santi1}. In that way, the classification of supersymmetric supergravity backgrounds is recovered in the Lie algebraic means. For different M-theory backgrounds in eleven dimensions, the deformation properties of Killing superalgebras are also classified in the literature \cite{OFarrill}.

In this paper, we consider KY and CKY superalgebras defined in constant curvature manifolds. After defining the relevant bilinear operations of these Lie superalgebras, the $\mathbb{Z}$-gradation and filtration properties of them are obtained. The dimensions of the superalgebras for different spacetime dimensions are determined. We investigate the cohomology theory of KY and CKY superalgebras by defining the spaces of cochains and the differential operations between them. By calculating the second cohomology group, we find the fact that both KY and CKY superalgebras are rigid and cannot be deformed to obtain new Lie superalgebras. This reveals the fact that they correspond to new geometric invariants of constant curvature manifolds. We also consider anti-de Sitter ($AdS$) spacetimes as examples of constant curvature manifolds, because of their importance in string theory and supergravity backgrounds. We discuss the structure of KY and CKY superalgebras in various dimensional $AdS$ spacetimes.

The paper is organized as follows. In Section 2, we define the SN bracket of KY superalgebras and their $\mathbb{Z}$-gradation and filtration properties. The cohomology theory and the proof of the rigidity of KY superalgebras are discussed in a subsection. Section 3 includes the definition of the CKY bracket and the $\mathbb{Z}$-gradation and filtration properties of CKY superalgebras with the proof of the rigidity of them. In Section 4, we consider the KY and CKY superalgebras of $AdS$ spacetimes as examples. Section 5 concludes the paper.

\section{KY superalgebras}

On an $n$-dimensional manifold $M$, we consider the space of differential forms $\Lambda M=\bigoplus_{p=1}^n\Lambda^p M$ where $\Lambda^p M$ denotes the space of $p$-forms on $M$. A $p$-form $\omega\in\Lambda^p M$ is called a KY $p$-form if it satisfies the following equation
\begin{equation}
\nabla_X\omega=\frac{1}{p+1}i_Xd\omega
\end{equation}
for any vector field $X$. Here $\nabla$ is the Levi-Civita connection and $\nabla_X$ denotes covariant derivative with respect to $X$. $i_X$ is the contraction or interior derivative operation with respect to $X$ and $d$ is the exterior derivative. For $p=1$, KY equation (1) reduces to the dual Killing equation that defines the 1-forms which are metric duals of the Killing vector fields. So, KY $p$-forms are antisymmetric generalizations of Killing vector fields to higher degree forms.

In constant curvature manifolds, KY forms satisfy a Lie superalgebra structure. A Lie superalgebra $\mathfrak{g}=\mathfrak{g}_{\bar{0}}\oplus\mathfrak{g}_{\bar{1}}$ is a direct sum of a Lie algebra $\mathfrak{g}_{\bar{0}}$ and a $\mathfrak{g}_{\bar{0}}$-module $\mathfrak{g}_{\bar{1}}$ together with a bilinear operation that has special symmetry properties. $\mathfrak{g}_{\bar{0}}$ is called the even part and $\mathfrak{g}_{\bar{1}}$ is called the odd part of the superalgebra. The bilinear operation $[.,.]$ on the Lie superalgebra $\mathfrak{g}$ is defined as
\[
[.,.]:\mathfrak{g}_i\times\mathfrak{g}_j\longrightarrow\mathfrak{g}_{i+j}
\]
for $i,j=\bar{0},\bar{1}(\text{mod }2)$. For $a,b,c\in\mathfrak{g}$, the bilinear operation $[.,.]$ satisfies the following skew-supersymmetry and super-Jacobi identities
\begin{eqnarray}
[a,b]&=&-(-1)^{|a||b|}[b,a]\nonumber\\
\left[a,[b,c]\right]&=&[[a, b], c]+(-1)^{|a||b|}[b, [a, c]]
\end{eqnarray}
where $|a|$ denotes the degree of $a$ that corresponds to 0 or 1 for $a$ is in $\mathfrak{g}_{\bar{0}}$ or $\mathfrak{g}_{\bar{1}}$, respectively.

On $\Lambda M$, the Schouten-Nijenhuis (SN) bracket which is defined for $\alpha\in\Lambda^p M$ and $\beta\in\Lambda^q M$ as
\begin{equation}
[\alpha,\beta]_{SN}=i_{X^a}\alpha\wedge\nabla_{X_a}\beta+(-1)^{pq}i_{X^a}\beta\wedge\nabla_{X_a}\alpha
\end{equation}
is a graded Lie bracket and gives a $(p+q-1)$-form. Here $\{X_a\}$ is an arbitrary basis of vector fields on $M$ and $\wedge$ denotes the wedge product on $\Lambda M$. It satisfies the following skew-symmetry identity
\begin{equation}
[\alpha, \beta]_{SN}=(-1)^{pq}[\beta, \alpha]_{SN}
\end{equation}
and the graded Jacobi identity
\begin{equation}
(-1)^{p(r+1)}[\alpha, [\beta, \gamma]_{SN}]_{SN}+(-1)^{q(p+1)}[\beta, [\gamma, \alpha]_{SN}]_{SN}+(-1)^{r(q+1)}[\gamma, [\alpha, \beta]_{SN}]_{SN}=0
\end{equation}
where $\gamma\in\Lambda^rM$.

We denote the space of KY $p$-forms on $M$ as $\Lambda^p_{KY}M$. For two KY forms $\omega_1\in\Lambda^p_{KY}M$ and $\omega_2\in\Lambda^q_{KY}M$, the SN bracket is written as
\begin{equation}
[\omega_1,\omega_2]_{SN}=\frac{1}{q+1}i_{X^a}\omega_1\wedge i_{X_a}d\omega_2+\frac{(-1)^{pq}}{p+1}i_{X^a}\omega_2\wedge i_{X_a}d\omega_1
\end{equation}
by using (1). In constant curvature manifolds, (6) gives a KY $(p+q-1)$-form and defines a bilinear operation on the space of KY forms \cite{Kastor Ray Traschen, Acik Ertem2}. Namely, we have
\begin{equation}
\nabla_X[\omega_1,\omega_2]_{SN}=\frac{1}{p+q}i_Xd[\omega_1,\omega_2]_{SN}.
\end{equation}
For $p=q=1$, the SN bracket in (6) reduces to the Lie bracket of Killing vector fields.

The SN bracket is a Lie bracket for odd degree KY forms since it takes two odd degree KY forms and gives another odd degree KY form. So, odd degree KY forms satisfy a Lie algebra structure with SN bracket. This gives way to the definition of the Lie superalgebra structure that is the KY superalgebra $\mathfrak{k}=\mathfrak{k}_{\bar{0}}\oplus\mathfrak{k}_{\bar{1}}$. The even part $\mathfrak{k}_{\bar{0}}$ of the superalgebra consists of the Lie algebra of odd degree KY forms and the odd part $\mathfrak{k}_{\bar{1}}$ of the superalgebra is the space of even degree KY forms. We have the following brackets of KY superalgebra; the SN bracket of two odd KY forms gives again an odd KY form:
\[
[ . , . ]_{SN}:\mathfrak{k}_{\bar{0}}\times\mathfrak{k}_{\bar{0}}\longrightarrow\mathfrak{k}_{\bar{0}}
\]
the SN bracket of an odd and an even KY forms gives an even KY form:
\[
[ . , . ]_{SN}:\mathfrak{k}_{\bar{0}}\times\mathfrak{k}_{\bar{1}}\longrightarrow\mathfrak{k}_{\bar{1}}
\]
and the SN bracket of two even KY forms gives an odd KY form:
\[
[ . , . ]_{SN}:\mathfrak{k}_{\bar{1}}\times\mathfrak{k}_{\bar{1}}\longrightarrow\mathfrak{k}_{\bar{0}}
\]
From (6), (4) and (5), one can see that the SN bracket of KY forms satisfy the skew-supersymmetry and super-Jacobi identities in (2). KY superalgebra is the extension of the Lie algebra structure of Killing vector fields which is a subalgebra in $\mathfrak{k}_{\bar{0}}$ to all KY forms in constant curvature manifolds.

$\mathbb{Z}$-graded structure of differential forms gives a $\mathbb{Z}$-gradation to the KY superalgebra. So, we can write in $n$-dimensions as
\[
\mathfrak{k}=\mathfrak{k}_0\oplus\mathfrak{k}_{-1}\oplus\mathfrak{k}_{-2}\oplus...\oplus\mathfrak{k}_{-(n-1)}=\bigoplus_{j=-(n-1)}^0\mathfrak{k}_j
\]
where we have defined the components as follows
\[
\mathfrak{k}_0=\Lambda^1_{KY}M\quad,\quad\mathfrak{k}_{-1}=\Lambda^2_{KY}M\quad,\quad\mathfrak{k}_{-2}=\Lambda^3_{KY}M\quad,\quad...\quad,\quad\mathfrak{k}_{-(n-1)}=\Lambda^n_{KY}M
\]
This means that we have a depth $(n-1)$ Lie superalgebra of KY forms in an $n$-dimensional constant curvature manifold $M$. The $\mathbb{Z}$-gradation we have defined is compatible with the $\mathbb{Z}_2$-gradation of the superalgebra and we have
\begin{eqnarray}
\mathfrak{k}_{\bar{0}}&=&\mathfrak{k}_0\oplus\mathfrak{k}_{-2}\oplus\mathfrak{k}_{-4}\oplus...\nonumber\\
\mathfrak{k}_{\bar{1}}&=&\mathfrak{k}_{-1}\oplus\mathfrak{k}_{-3}\oplus\mathfrak{k}_{-5}\oplus...\nonumber
\end{eqnarray}
The $\mathbb{Z}$-graded structure of KY superalgebra gives also a natural filtration $F$ to it such as
\[
F=F_{-(n-1)}\supset F_{-(n-2)}\supset...\supset F_{-2}\supset F_{-1}\supset F_0
\]
and the filtration components are defined as
\begin{eqnarray}
F_0&=&\mathfrak{k}_0\nonumber\\
F_{-1}&=&\mathfrak{k}_0\oplus\mathfrak{k}_{-1}\nonumber\\
F_{-2}&=&\mathfrak{k}_0\oplus\mathfrak{k}_{-1}\oplus\mathfrak{k}_{-2}\nonumber\\
&&...\nonumber\\
F_{-(n-2)}&=&\mathfrak{k}_0\oplus\mathfrak{k}_{-1}\oplus...\oplus\mathfrak{k}_{-(n-2)}\nonumber\\
F_{-(n-1)}&=&\mathfrak{k}\nonumber
\end{eqnarray}

Let us denote the dimension of a Lie superalgebra as $(\alpha|\beta)$ where $\alpha$ corresponds to the dimension of the even part of the superalgebra and $\beta$ is the dimension of the odd part. In constant curvature manifolds, the number of KY forms is maximal and it can be calculated by using the following integrability condition of the KY equation \cite{Acik Ertem Onder Vercin2}
\begin{equation}
\nabla_{X_b}d\omega=\frac{p+1}{p}R_{ab}\wedge i_{X^a}\omega
\end{equation}
where $R_{ab}$ denotes the curvature 2-forms. Since it determines the second and higher order derivatives of KY forms in terms of themselves, the counting gives the maximal number of KY $p$-forms in $n$-dimensions as follows \cite{Kastor Ray Traschen}
\begin{equation}
K_p=\left(
             \begin{array}{c}
               n+1 \\
               p+1 \\
             \end{array}
           \right)=\frac{(n+1)!}{(p+1)!(n-p)!}.
\end{equation}
So, the dimension of the KY superalgebra is $(K_{\text{odd}}|K_{\text{even}})$ and $K_{\text{odd}}$ and $K_{\text{even}}$ are the number of odd and even degree KY forms respectively which are defined as follows
\begin{equation}
K_{\text{odd}}=\sum_{k=1}^{\lfloor\frac{n}{2}\rfloor}\left(
             \begin{array}{c}
               n+1 \\
               2k \\
             \end{array}
           \right)
\end{equation}
\begin{equation}
K_{\text{even}}=\sum_{k=1}^{\lfloor\frac{n-1}{2}\rfloor}\left(
             \begin{array}{c}
               n+1 \\
               2k+1 \\
             \end{array}
           \right)
\end{equation}
where $\lfloor\,\rfloor$ denotes the floor function which takes the integer part of the argument.

\subsection{Cohomology and rigidity}

We consider the cohomology theory of Lie superalgebras to investigate the possible deformations of KY superalgebras. To do this, we first define the spaces of cochains in the superalgebra. In a Lie superalgebra $\mathfrak{g}$, the space of $p$-cochains which takes values in $\mathfrak{g}$ is denoted by $C^p(\mathfrak{g};\mathfrak{g})$ and is defined as
\[
C^p(\mathfrak{g};\mathfrak{g}):=\left\{f:\underbrace{\mathfrak{g}\times\mathfrak{g}\times ...\times\mathfrak{g}}_{p \text{ times}}\longrightarrow\mathfrak{g}\, \bigg| \,f(a_1,...,a_p)=-(-1)^{|a_i||a_{i+1}|}f(a_1,...,a_{i-1},a_{i+1},a_i,a_{i+2},...,a_p) \right\}
\]
where $a_i\in\mathfrak{g}$. So, the elements of $C^p(\mathfrak{g};\mathfrak{g})$ are skewsymmetric $p$-linear maps in $\mathfrak{g}$. However, in a filtered $\mathbb{Z}$-graded Lie superalgebra, the space of $p$-cochains can be refined by using the filtration degree of the maps. The filtration degree of a $p$-cochain
\[
f:\mathfrak{g}_{i_1}\times\mathfrak{g}_{i_2}\times ...\times\mathfrak{g}_{i_p}\longrightarrow\mathfrak{g}_j
\]
is defined as $d=j-(i_1+i_2+...+i_p)$ where $i_1,...i_p,j$ denotes the $\mathbb{Z}$-gradation degrees in $\mathfrak{g}$. Then, $C^p(\mathfrak{g};\mathfrak{g})$ can be written as a sum of spaces of $p$-cochains with different filtration degrees
\[
C^{d,p}(\mathfrak{g};\mathfrak{g}):=\big\{f:\mathfrak{g}_{i_1}\times\mathfrak{g}_{i_2}\times ...\times\mathfrak{g}_{i_p}\longrightarrow\mathfrak{g}_j\, \big| \,f \text{ is a $p$-cochain with } d=j-(i_1+i_2+...+i_p) \big\}
\]
\[
C^p({\mathfrak{g};\mathfrak{g}})=\bigoplus_d C^{d,p}(\mathfrak{g};\mathfrak{g}).
\]

We define the differential $d:C^{d,p}(\mathfrak{g};\mathfrak{g})\longrightarrow C^{d,p+1}(\mathfrak{g};\mathfrak{g})$ as follows. For $f\in C^{d,p}(\mathfrak{g};\mathfrak{g})$
\begin{eqnarray}
df(a_0,a_1,...,a_p)&=&\sum_{i=0}^p(-1)^{i+|a_i|(|f|+|a_0|+...+|a_{i-1}|)}a_i.f(a_0,...,\widehat{a_i},...,a_p)\\
&+&\sum_{0\leq i<j\leq p}(-1)^{i+j+(|a_i|+|a_j|)(|a_0|+...+|a_{i-1}|)+|a_j|(|a_{i+1}|+...+|a_{j-1}|)}f([a_i,a_j],a_0,...,\widehat{a_i},...,\widehat{a_j},...,a_p)\nonumber
\end{eqnarray}
where $.$ denotes the action of the superalgebra on itself and $\,\widehat{  }\,$ indicates the omitting of the element below it. The differential obeys the identity $d^2=0$ and we can define $p$-cocycles as $\text{ker }d:C^{d,p}\longrightarrow C^{d,p+1}$ and $p$-coboundaries as $\text{im } d:C^{d,p-1}\longrightarrow C^{d,p}$ from $p$-cochains by using it. This gives rise to the following cohomology groups of Lie superalgebras
\[
H^{d,p}(\mathfrak{g};\mathfrak{g}):=\frac{\text{ker }d:C^{d,p}\longrightarrow C^{d,p+1}}{\text{im } d:C^{d,p-1}\longrightarrow C^{d,p}}
\]
and the defined cohomology is called the generalized Spencer cohomology of filtered $\mathbb{Z}$-graded Lie superalgebras \cite{Cheng Kac}. Consequently, the $p$-th cohomology group of the Lie superalgebra is defined as
\[
H^p({\mathfrak{g};\mathfrak{g}})=\bigoplus_d H^{d,p}(\mathfrak{g};\mathfrak{g}).
\]

A deformation of a Lie superalgebra corresponds to a one-parameter family of bilinear operations $[.,.]_{\epsilon}$ on $\mathfrak{g}$ and for $\epsilon=0$ it is equivalent to the original bracket operation $[.,.]$ on $\mathfrak{g}$. So, for $a,b\in\mathfrak{g}$, the deformed bracket can be written as follows
\begin{equation}
[a,b]_{\epsilon}=[a,b]+\epsilon P_1(a,b)+\epsilon^2P_2(a,b)+...=\sum_{i\geq 0}\epsilon^iP_i(a,b)
\end{equation}
where $[a,b]_0=[a,b]$. The conditions of skew-supersymmetry and graded Jacobi identity on the deformed bracket gives various restrictions on all deformation functions $P_i$. If we have one deformation function $P_1$, then it is called the infinitesimal deformation of the bracket and the restriction on $P_1$ corresponds to a cocycle condition on the space of 2-cochains $C^2(\mathfrak{g};\mathfrak{g})$ \cite{OFarrill}. Indeed, the nontrivial infinitesimal deformations are equivalent to the nontrivial elements of the cohomology group $H^2(\mathfrak{g};\mathfrak{g})$. If $H^2(\mathfrak{g};\mathfrak{g})=0$, then the Lie superalgebra cannot be deformed and it is called as a rigid Lie superalgebra. The higher cohomology groups give the obstructions to the integrability of the infinitesimal deformation.

Now, we demonstrate the spaces of cochains in the KY superalgebra. It can easily be seen from the definition of the SN bracket in (3) that the SN bracket of a 1-form with a $p$-form always gives a $p$-form. So, SN bracket with respect to a 1-form does not change the degree of the forms. This means that 0-cochains of KY superalgebra consists of the space of KY 1-forms $\mathfrak{k}_0=\Lambda^1_{KY}M$
\[
C^0(\mathfrak{k};\mathfrak{k})=\left\{\phi:\mathfrak{k}_0\right\}.
\]
1-cochains correspond to the multiples of identity maps between the same degree KY forms. Hence, we have
\[
C^1(\mathfrak{k};\mathfrak{k})=\left\{\phi_i:\mathfrak{k}_i\longrightarrow\mathfrak{k}_i\,\big|\, i=0,-1,-2,...,-(n-1)\right\}.
\]
Since all the filtration degrees of 1-cochains are zero, one obtains $C^1(\mathfrak{k};\mathfrak{k})=C^{0,1}(\mathfrak{k};\mathfrak{k})$. Space of 2-cochains is determined by the defined bilinear operation of the Lie superalgebra. Then, SN bracket determines the 2-cochains of KY superalgebra and because of the fact that SN bracket produces a $(p+q-1)$-form from a $p$-form and a $q$-form, together with the definition $\Lambda^p_{KY}M=\mathfrak{k}_{-(p-1)}$, the space of 2-cochains is given by
\[
C^2(\mathfrak{k};\mathfrak{k})=\left\{\phi_{ij}:\mathfrak{k}_i\times\mathfrak{k}_j\longrightarrow\mathfrak{k}_{i+j}\,\big|\, i,j=0,-1,-2,...,-(n-1)\right\}.
\]
Again, it can easily be seen that the filtration degrees of 2-cochains are zero and we have $C^2(\mathfrak{k};\mathfrak{k})=C^{0,2}(\mathfrak{k};\mathfrak{k})$. The 3-cochains correspond to applying the bilinear operation two times successively and from the properties of the SN bracket we obtain
\[
C^3(\mathfrak{k};\mathfrak{k})=\left\{\phi_{ijk}:\mathfrak{k}_i\times\mathfrak{k}_j\times\mathfrak{k}_k\longrightarrow\mathfrak{k}_{i+j+k}\,\big|\, i,j,k=0,-1,-2,...,-(n-1)\right\}.
\]
Filtration degrees are again zero and $C^3(\mathfrak{k};\mathfrak{k})=C^{0,3}(\mathfrak{k};\mathfrak{k})$. The higher cochain spaces can be determined in a similar way. However, to obtain the second cohomology group of the KY superalgebra, which gives the possible deformations, it will be enough to know the spaces of 1-, 2- and 3-cochains.

Let us define the differential operators $d$ between the spaces of $p$-cochains for the KY superalgebra;

i) For $d:C^0(\mathfrak{k};\mathfrak{k})\longrightarrow C^1(\mathfrak{k};\mathfrak{k})$, by taking $\phi=\mathfrak{k}_0\in C^0(\mathfrak{k};\mathfrak{k})$ and $a\in\mathfrak{k}_i$ for $i=0,-1,...,-(n-1)$, we can write the differential from the definition (12) as
\begin{equation}
d\phi(a)=[a,\phi]_{SN}.
\end{equation}
Since $\phi$ is a 1-form, the SN bracket does not change the degree of $a$ and we have $d\phi:\mathfrak{k}_i\longrightarrow\mathfrak{k}_i\,\in\,C^1(\mathfrak{k};\mathfrak{k})$.

ii) For $d:C^1(\mathfrak{k};\mathfrak{k})\longrightarrow C^2(\mathfrak{k};\mathfrak{k})$, we take $a\in\mathfrak{k}_i$, $b\in\mathfrak{k}_j$ and $\phi_i,\phi_j,\phi_{i+j}\in C^1(\mathfrak{k};\mathfrak{k})$, then the differential is
\begin{equation}
d\phi_{i,j}(a,b)=[a,\phi_j(b)]_{SN}-(-1)^{|a||b|}[b,\phi_i(a)]_{SN}-\phi_{i+j}([a,b]_{SN}).
\end{equation}
So, we have $d\phi_{i,j}:\mathfrak{k}_i\times\mathfrak{k}_j\longrightarrow\mathfrak{k}_{i+j}\,\in\,C^2(\mathfrak{k};\mathfrak{k})$.

iii) For $d:C^2(\mathfrak{k};\mathfrak{k})\longrightarrow C^3(\mathfrak{k};\mathfrak{k})$, from the elements $a\in\mathfrak{k}_i$, $b\in\mathfrak{k}_j$, $c\in\mathfrak{k}_k$ and $\phi_{ij},\phi_{jk},\phi_{ki},\phi_{i+j,k},\phi_{j+k,i},\phi_{k+i,j}\in C^2(\mathfrak{k},\mathfrak{k})$, the differential is obtained as
\begin{eqnarray}
d\phi_{i,j,k}(a,b,c)&=&[a,\phi_{jk}(b,c)]_{SN}+(-1)^{|a|(|b|+|c|)}[b,\phi_{ki}(c,a)]_{SN}+(-1)^{|c|(|a|+|b|)}[c,\phi_{ij}(a,b)]_{SN}\nonumber\\
&&-\phi_{i+j,k}([a,b]_{SN},c)-(-1)^{|a|(|b|+|c|)}\phi_{j+k,i}([b,c]_{SN},a)-(-1)^{|c|(|a|+|b|)}\phi_{k+i,j}([c,a]_{SN},b)
\end{eqnarray}
and we have $d\phi_{i,j,k}:\mathfrak{k}_i\times\mathfrak{k}_j\times\mathfrak{k}_k\longrightarrow\mathfrak{k}_{i+j+k}\,\in\,C^3(\mathfrak{k};\mathfrak{k})$.

Since the deformations of a Lie superalgebra is characterized by the cohomology group $H^2(\mathfrak{k};\mathfrak{k})$, we will analyze the cocycle and coboundary conditions of the elements in $C^2(\mathfrak{k};\mathfrak{k})$. Let us start with the map $d:C^1(\mathfrak{k};\mathfrak{k})\longrightarrow C^2(\mathfrak{k};\mathfrak{k})$ and the elements $a\in\mathfrak{k}_i$ and $b\in\mathfrak{k}_j$. We know that all $\phi_i\in C^1(\mathfrak{k};\mathfrak{k})$ are identity maps; $\phi_i(a)=a$ and $\phi_{ij}\in C^2(\mathfrak{k};\mathfrak{k})$ correspond to SN brackets; $\phi_{ij}(a,b)=[a,b]_{SN}$. If we calculate the differential of $\phi_i$, we find
\begin{eqnarray}
d\phi_{i,j}(a,b)&=&[a,\phi_j(b)]_{SN}-(-1)^{|a||b|}[b,\phi_i(a)]_{SN}-\phi_{i+j}([a,b]_{SN})\nonumber\\
&=&[a,b]_{SN}-(-1)^{|a||b|}[b,a]_{SN}-[a,b]_{SN}\nonumber\\
&=&-(-1)^{|a||b|}[b,a]_{SN}\nonumber\\
&=&[a,b]_{SN}.\nonumber
\end{eqnarray}
This means that we have
\begin{equation}
d\phi_{i,j}(a,b)=\phi_{ij}(a,b)
\end{equation}
for all $i,j$. Then, all 2-cochains in $C^2(\mathfrak{k};\mathfrak{k})$ are coboundaries. For the map $d:C^2(\mathfrak{k};\mathfrak{k})\longrightarrow C^3(\mathfrak{k};\mathfrak{k})$ and the elements $a\in\mathfrak{k}_i$, $b\in\mathfrak{k}_j$ and $c\in\mathfrak{k}_k$, we calculate the differential $d\phi_{i,j,k}$ of the elements $\phi_{ij}\in C^2(\mathfrak{k};\mathfrak{k})$ as
\begin{eqnarray}
d\phi_{i,j,k}(a,b,c)&=&[a,\phi_{jk}(b,c)]_{SN}+(-1)^{|a|(|b|+|c|)}[b,\phi_{ki}(c,a)]+(-1)^{|c|(|a|+|b|)}[c,\phi_{ij}(a,b)]_{SN}\nonumber\\
&&-\phi_{i+j,k}([a,b]_{SN},c)-(-1)^{|a|(|b|+|c|)}\phi_{j+k,i}([b,c]_{SN},a)-(-1)^{|c|(|a|+|b|)}\phi_{k+i,j}([c,a]_{SN},b)\nonumber\\
&=&[a,[b,c]_{SN}]_{SN}+(-1)^{|a|(|b|+|c|)}[b,[c,a]_{SN}]_{SN}+(-1)^{|c|(|a|+|b|)}[c,[a,b]_{SN}]_{SN}\nonumber\\
&&-[[a,b]_{SN},c]_{SN}-(-1)^{|a|(|b|+|c|)}[[b,c]_{SN},a]_{SN}-(-1)^{|c|(|a|+|b|)}[[c,a]_{SN},b]_{SN}.\nonumber
\end{eqnarray}
By using the following identities
\[
(-1)^{|a|(|b|+|c|)}[b,[c,a]_{SN}]_{SN}=-(-1)^{|a|b|}[b,[a,c]_{SN}]_{SN}
\]
\[
-(-1)^{|a|(|b|+|c|)}[[b,c]_{SN},a]_{SN}=-(-1)^{|b||c|}[a,[c,b]_{SN}]_{SN}
\]
we obtain
\begin{eqnarray}
d\phi_{i,j,k}(a,b,c)&=&[a,[b,c]_{SN}]_{SN}-[[a,b]_{SN},c]_{SN}-(-1)^{|a||b|}[b,[c,a]_{SN}]_{SN}\nonumber\\
&&+(-1)^{|c|(|a|+|b|)}\left([c,[a,b]_{SN}]_{SN}-[[c,a]_{SN},b]_{SN}-(-1)^{|c||a|}[a,[c,b]_{SN}]_{SN}\right)\nonumber
\end{eqnarray}
and from the super-Jacobi identities
\[
[a,[b,c]_{SN}]_{SN}-[[a,b]_{SN},c]_{SN}-(-1)^{|a||b|}[b,[c,a]_{SN}]_{SN}=0
\]
\[
[c,[a,b]_{SN}]_{SN}-[[c,a]_{SN},b]_{SN}-(-1)^{|c||a|}[a,[c,b]_{SN}]_{SN}=0
\]
we have
\begin{equation}
d\phi_{i,j,k}(a,b,c)=0
\end{equation}
for all $i,j,k$. This means that all 2-cochains in $C^2(\mathfrak{k};\mathfrak{k})$ are cocycles. So, from the previous analysis, we obtain the following identities
\begin{eqnarray}
\text{im }d:C^1(\mathfrak{k};\mathfrak{k})\longrightarrow C^2(\mathfrak{k};\mathfrak{k})&=&C^2(\mathfrak{k};\mathfrak{k})\nonumber\\
\text{ker }d:C^2(\mathfrak{k};\mathfrak{k})\longrightarrow C^3(\mathfrak{k};\mathfrak{k})&=&C^2(\mathfrak{k};\mathfrak{k})\nonumber
\end{eqnarray}
and this gives the triviality of the second cohomology group as follows
\begin{equation}
H^2(\mathfrak{k};\mathfrak{k})=\frac{\text{ker }d:C^2(\mathfrak{k};\mathfrak{k})\longrightarrow C^3(\mathfrak{k};\mathfrak{k})}{\text{im }d:C^1(\mathfrak{k};\mathfrak{k})\longrightarrow C^2(\mathfrak{k};\mathfrak{k})}=\frac{C^2(\mathfrak{k};\mathfrak{k})}{C^2(\mathfrak{k};\mathfrak{k})}=0.
\end{equation}
Then, the KY superalgebras of constant curvature manifolds are rigid Lie superalgebras whose bilinear bracket cannot be deformed to obtain other Lie superalgebras. This reveals the fact that KY superalgebras correspond to geometric invariants for constant curvature manifolds.

\section{CKY superalgebras}

KY forms are special cases of more general conformal structures that are called CKY forms on a manifold $M$. In $n$ dimensions, a $p$-form $\omega\in\Lambda^pM$ is called a CKY $p$-form if it satisfies the equation
\begin{equation}
\nabla_X\omega=\frac{1}{p+1}i_Xd\omega-\frac{1}{n-p+1}\widetilde{X}\wedge\delta\omega
\end{equation} 
for any vector field $X$, where $\widetilde{X}$ denotes the 1-form corresponding to the metric dual of $X$. For $p=1$, CKY equation reduces to the dual conformal Killing equation whose solutions define the 1-forms that are metric duals of the conformal Killing vector fields. Moreover, it reduces to the KY equation for the CKY forms that satisfy the condition $\delta\omega=0$, where $\delta$ denotes the coderivative operation. So, KY forms constitute a subset in the space of CKY forms and they correspond to the co-closed CKY forms, namely $\delta\omega=0$.

In constant curvature manifolds, CKY forms also satisfy a Lie superalgebra structure. However, the bilinear operation of this Lie superalgebra does not correspond to the SN bracket defined in the previous section. We denote the space of CKY $p$-forms on $M$ as $\Lambda^p_{CKY}M$. For two CKY forms $\omega_1\in\Lambda^p_{CKY}M$ and $\omega_2\in\Lambda^q_{CKY}M$, let us define the CKY bracket as follows
\begin{eqnarray}
[\omega_1, \omega_2]_{CKY}&=&\frac{1}{q+1}i_{X_a}\omega_1\wedge i_{X^a}d\omega_2+\frac{(-1)^p}{p+1}i_{X_a}d\omega_1\wedge i_{X^a}\omega_2\nonumber\\
&&+\frac{(-1)^p}{n-q+1}\omega_1\wedge\delta\omega_2+\frac{1}{n-p+1}\delta\omega_1\wedge\omega_2.
\end{eqnarray}
This bracket gives a ($p+q-1$)-form and it satifies the CKY equation \cite{Ertem4}, namely we have
\begin{equation}
\nabla_{X_a}[\omega_1, \omega_2]_{CKY}=\frac{1}{p+q}i_{X_a}d[\omega_1, \omega_2]_{CKY}-\frac{1}{n-p-q+2}e_a\wedge\delta[\omega_1, \omega_2]_{CKY}.
\end{equation}
For $p=q=1$, the CKY bracket reduces to the Lie bracket of vector fields. Moreover, for the subset of KY forms that is $\delta\omega_1=0$ and $\delta\omega_2=0$, it reduces to the SN bracket of KY forms. However, for general CKY forms, the CKY bracket differs from the SN bracket in terms of the coefficients of the third and fourth terms in (21). It can also be seen from the definition that the CKY bracket has the following skew-supersymmetry and graded Jacobi identity properties
\begin{eqnarray}
[\alpha, \beta]_{CKY}&=&(-1)^{pq}[\beta, \alpha]_{CKY}\\
(-1)^{p(r+1)}[\alpha, [\beta, \gamma]_{CKY}]_{CKY}&+&(-1)^{q(p+1)}[\beta, [\gamma, \alpha]_{CKY}]_{CKY}+(-1)^{r(q+1)}[\gamma, [\alpha, \beta]_{CKY}]_{CKY}=0
\end{eqnarray}
for any $\alpha\in\Lambda^pM$, $\beta\in\Lambda^qM$ and $\gamma\in\Lambda^rM$. So, it is a graded Lie bracket for CKY forms.

Moreover, a subset of CKY forms that are called normal CKY forms also satisfy a Lie superalgebra structure in Einstein manifolds. The normal CKY forms satisfy the following integrability condition besides the CKY equation defined in (20)
\begin{equation}
\frac{p}{p+1}\delta d\omega+\frac{n-p}{n-p+1}d\delta\omega=-2(n-p)K_a\wedge i_{X^a}\omega
\end{equation}
where the 1-form $K_a$ is defined in terms of the Ricci 1-forms $P_a$, the curvature scalar ${\cal{R}}$ and the orthonormal coframe basis $e_a$ as
\begin{equation}
K_a=\frac{1}{n-2}\left(\frac{\cal{R}}{2(n-1)}e_a-P_a\right).
\end{equation}
The CKY bracket of two normal CKY forms gives another normal CKY form and it defines a graded Lie bracket for normal CKY forms in Einstein manifolds \cite{Ertem4}.

Odd degree CKY forms constitute a Lie algebra under the CKY bracket since it takes two odd degree CKY forms and gives another odd degree CKY form. So, we can define a Lie superalgebra structure that is the CKY superalgebra $\mathfrak{c}=\mathfrak{c}_{\bar{0}}\oplus\mathfrak{c}_{\bar{1}}$. The even part $\mathfrak{c}_{\bar{0}}$ consists of the Lie algebra of odd degree CKY forms and the odd part $\mathfrak{c}_{\bar{1}}$ is the space of even degree CKY forms. The brackets of CKY superalgebra are as follows; the CKY bracket of two odd CKY forms gives again an odd CKY form:
\[
[ . , . ]_{CKY}:\mathfrak{c}_{\bar{0}}\times\mathfrak{c}_{\bar{0}}\longrightarrow\mathfrak{c}_{\bar{0}}
\]
the CKY bracket of an odd and an even CKY forms gives an even CKY form:
\[
[ . , . ]_{CKY}:\mathfrak{c}_{\bar{0}}\times\mathfrak{c}_{\bar{1}}\longrightarrow\mathfrak{c}_{\bar{1}}
\]
and the CKY bracket of two even CKY forms gives an odd CKY form:
\[
[ . , . ]_{CKY}:\mathfrak{c}_{\bar{1}}\times\mathfrak{c}_{\bar{1}}\longrightarrow\mathfrak{c}_{\bar{0}}
\]
CKY superalgebra is the extension of the Lie algebra of conformal Killing vector fields which is a subalgebra in $\mathfrak{c}_{\bar{0}}$ to all CKY forms in constant curvature manifolds. It is also an extension of the KY superalgebra since $\mathfrak{k}_{\bar{0}}\subset\mathfrak{c}_{\bar{0}}$ and $\mathfrak{k}_{\bar{1}}\subset\mathfrak{c}_{\bar{1}}$. So, KY superalgebra is a subsuperalgebra of the CKY superalgebra.

$\mathbb{Z}$-graded structure of CKY superalgebra can be defined similarly to the case of the KY superalgebra. In $n$-dimensions we have
\[
\mathfrak{c}=\mathfrak{c}_0\oplus\mathfrak{c}_{-1}\oplus\mathfrak{c}_{-2}\oplus...\oplus\mathfrak{c}_{-(n-1)}=\bigoplus_{j=-(n-1)}^0\mathfrak{c}_j
\]
where the $\mathbb{Z}$-gradation degrees are defined as follows
\[
\mathfrak{c}_0=\Lambda^1_{CKY}M\quad,\quad\mathfrak{c}_{-1}=\Lambda^2_{CKY}M\quad,\quad\mathfrak{c}_{-2}=\Lambda^3_{CKY}M\quad,\quad...\quad,\quad\mathfrak{c}_{-(n-1)}=\Lambda^n_{CKY}M.
\]
This means that we have a depth $(n-1)$ Lie superalgebra of CKY forms in an $n$-dimensional constant curvature manifold $M$. This $\mathbb{Z}$-gradation is compatible with the $\mathbb{Z}_2$-gradation of the superalgebra and we have
\begin{eqnarray}
\mathfrak{c}_{\bar{0}}&=&\mathfrak{c}_0\oplus\mathfrak{c}_{-2}\oplus\mathfrak{c}_{-4}\oplus...\nonumber\\
\mathfrak{c}_{\bar{1}}&=&\mathfrak{c}_{-1}\oplus\mathfrak{c}_{-3}\oplus\mathfrak{c}_{-5}\oplus...\nonumber
\end{eqnarray}
Moreover, the filtration components of the CKY superalgebra are defined as
\begin{eqnarray}
F_0&=&\mathfrak{c}_0\nonumber\\
F_{-1}&=&\mathfrak{c}_0\oplus\mathfrak{c}_{-1}\nonumber\\
F_{-2}&=&\mathfrak{c}_0\oplus\mathfrak{c}_{-1}\oplus\mathfrak{c}_{-2}\nonumber\\
&&...\nonumber\\
F_{-(n-2)}&=&\mathfrak{c}_0\oplus\mathfrak{c}_{-1}\oplus...\oplus\mathfrak{c}_{-(n-2)}\nonumber\\
F_{-(n-1)}&=&\mathfrak{c}\nonumber
\end{eqnarray}

The dimension of the CKY superalgebra is maximal in constant curvature manifolds since the number of CKY forms is maximal in that case. The following integrability condition of CKY equation can be used to determine the number of CKY forms
\begin{equation}
\frac{p}{p+1}\delta d\omega+\frac{n-p}{n-p+1}d\delta\omega=P_a\wedge i_{X^a}\omega+R_{ab}\wedge i_{X^a}i_{X^b}\omega.
\end{equation}
It determines the second and higher order derivatives of CKY forms in terms of themselves and counting the maximal number of CKY $p$-forms in $n$-dimensions gives \cite{Semmelmann}
\begin{equation}
C_p=\left(
             \begin{array}{c}
               n+2 \\
               p+1 \\
             \end{array}
           \right)=\frac{(n+2)!}{(p+1)!(n-p+1)!}.
\end{equation}
So, the dimension of the CKY superalgebra is $(C_{\text{odd}}|C_{\text{even}})$ and $C_{\text{odd}}$ and $C_{\text{even}}$ are the number of odd and even degree CKY forms respectively which are defined as follows
\begin{equation}
C_{\text{odd}}=\sum_{k=1}^{\lfloor\frac{n}{2}\rfloor}\left(
             \begin{array}{c}
               n+2 \\
               2k \\
             \end{array}
           \right)
\end{equation}
\begin{equation}
C_{\text{even}}=\sum_{k=1}^{\lfloor\frac{n-1}{2}\rfloor}\left(
             \begin{array}{c}
               n+2 \\
               2k+1 \\
             \end{array}
           \right).
\end{equation}
These differs from the KY superalgebra case by taking $n+2$ in the combinations instead of $n+1$. All KY forms are CKY forms at the same time and the number of CKY $p$-forms that do not correspond to KY $p$-forms is given by
\begin{eqnarray}
C_p-K_p=\left(
             \begin{array}{c}
               n+2 \\
               p+1 \\
             \end{array}
           \right)-\left(
             \begin{array}{c}
               n+1 \\
               p+1 \\
             \end{array}
           \right)=\left(
             \begin{array}{c}
               n+1 \\
               p \\
             \end{array}\right).
\end{eqnarray}

The spaces of cochains in the CKY superalgebra are similar to the KY superalgebra case. The CKY bracket of a 1-form with a $p$-form gives again a $p$-form and 0-cochains of CKY superalgebra corresponds to the space of CKY 1-forms $\mathfrak{c}_0=\Lambda^1_{CKY}M$
\[
C^0(\mathfrak{c};\mathfrak{c})=\left\{\phi:\mathfrak{c}_0\right\}.
\]
1-cochains are multiples of identity maps between the same degree CKY forms
\[
C^1(\mathfrak{c};\mathfrak{c})=\left\{\phi_i:\mathfrak{c}_i\longrightarrow\mathfrak{c}_i\,\big|\,i=0,-1,-2,...,-(n-1)\right\}.
\]
All the filtration degrees of 1-cochains are zero and we have $C^1(\mathfrak{c};\mathfrak{c})=C^{0,1}(\mathfrak{c};\mathfrak{c})$. The space of 2-cochains are determined by the CKY bracket and it is given by
\[
C^2(\mathfrak{c};\mathfrak{c})=\left\{\phi_{ij}:\mathfrak{c}_i\times\mathfrak{c}_j\longrightarrow\mathfrak{c}_{i+j}\,\big|\,i,j=0,-1,-2,...,-(n-1)\right\}.
\]
Since the filtration degrees of 2-cochains are zero, we again have $C^2(\mathfrak{c};\mathfrak{c})=C^{0,2}(\mathfrak{c};\mathfrak{c})$. 3-cochains are obtained by applying the bilinear operation two times successively and given by
\[
C^3(\mathfrak{c};\mathfrak{c})=\left\{\phi_{ijk}:\mathfrak{c}_i\times\mathfrak{c}_j\times\mathfrak{c}_k\longrightarrow\mathfrak{c}_{i+j+k}\,\big|\,i,j,k=0,-1,-2,...,-(n-1)\right\}.
\]
The filtration degrees are again zero and $C^3(\mathfrak{c};\mathfrak{c})=C^{0,3}(\mathfrak{c};\mathfrak{c})$.

The differential operators $d$ between $p$-cochains are defined from the CKY bracket;

i) For $\phi=\mathfrak{c}_0\in C^0(\mathfrak{c};\mathfrak{c})$ and $a\in\mathfrak{c}_i$, $i=0,-1,...,-(n-1)$, the operator $d:C^0(\mathfrak{c};\mathfrak{c})\longrightarrow C^1(\mathfrak{c};\mathfrak{c})$ is written as
\begin{equation}
d\phi(a)=[a,\phi]_{CKY}
\end{equation}
and we have $d\phi:\mathfrak{c}_i\longrightarrow\mathfrak{c}_i\in C^1(\mathfrak{c};\mathfrak{c})$.

ii) For $a\in\mathfrak{c}_i$, $b\in\mathfrak{c}_j$ and $\phi_i,\phi_j,\phi_{i+j}\in C^1(\mathfrak{c};\mathfrak{c})$, the differential $d:C^1(\mathfrak{c};\mathfrak{c})\longrightarrow C^2(\mathfrak{c};\mathfrak{c})$ is
\begin{equation}
d\phi_{i,j}(a,b)=[a,\phi_j(b)]_{CKY}-(-1)^{|a||b|}[b,\phi_i(a)]_{CKY}-\phi_{i+j}([a,b]_{CKY})
\end{equation}
and $d\phi_{i,j}:\mathfrak{c}_i\times\mathfrak{c}_j\longrightarrow\mathfrak{c}_{i+j}\,\in\,C^2(\mathfrak{c};\mathfrak{c})$.

iii) For $a\in\mathfrak{c}_i$, $b\in\mathfrak{c}_j$, $c\in\mathfrak{c}_k$ and $\phi_{ij},\phi_{jk},\phi_{ki},\phi_{i+j,k},\phi_{j+k,i},\phi_{k+i,j}\in C^2(\mathfrak{c},\mathfrak{c})$, the operator $d:C^2(\mathfrak{c};\mathfrak{c})\longrightarrow C^3(\mathfrak{c};\mathfrak{c})$ is
\begin{eqnarray}
d\phi_{i,j,k}(a,b,c)&=&[a,\phi_{jk}(b,c)]_{CKY}+(-1)^{|a|(|b|+|c|)}[b,\phi_{ki}(c,a)]_{CKY}+(-1)^{|c|(|a|+|b|)}[c,\phi_{ij}(a,b)]_{CKY}\\
&&-\phi_{i+j,k}([a,b]_{CKY},c)-(-1)^{|a|(|b|+|c|)}\phi_{j+k,i}([b,c]_{CKY},a)-(-1)^{|c|(|a|+|b|)}\phi_{k+i,j}([c,a]_{CKY},b)\nonumber
\end{eqnarray}
and we have $d\phi_{i,j,k}:\mathfrak{c}_i\times\mathfrak{c}_j\times\mathfrak{c}_k\longrightarrow\mathfrak{c}_{i+j+k}\,\in\,C^3(\mathfrak{c};\mathfrak{c})$.

Since all $\phi_i\in C^1(\mathfrak{c};\mathfrak{c})$ are identity maps $\phi_i(a)=a$ and $\phi_{ij}(a,b)=[a,b]_{CKY}$, the differentail of $\phi_i$ gives
\[
d\phi_{i,j}(a,b)=[a,b]_{CKY}=\phi_{ij}(a,b)
\]
for all $i,j$. So, all 2-cochains in $C^2(\mathfrak{c};\mathfrak{c})$ are coboundaries. Similar calculations for the differential $d\phi_{i,j,k}$ of the elements $\phi_{ij}\in C^2(\mathfrak{c};\mathfrak{c})$ can be done as in the KY superalgebra case by using the properties of the CKY bracket and one obtains
\[
d\phi_{i,j,k}(a,b,c)=0
\]
for all $i,j,k$. This shows that all 2-cochains in $C^2(\mathfrak{c};\mathfrak{c})$ are cocycles. Then we have
\begin{eqnarray}
\text{im }d:C^1(\mathfrak{c};\mathfrak{c})\longrightarrow C^2(\mathfrak{c};\mathfrak{c})&=&C^2(\mathfrak{c};\mathfrak{c})\nonumber\\
\text{ker }d:C^2(\mathfrak{c};\mathfrak{c})\longrightarrow C^3(\mathfrak{c};\mathfrak{c})&=&C^2(\mathfrak{c};\mathfrak{c})\nonumber
\end{eqnarray}
and this corresponds to the triviality of the second cohomology group
\begin{equation}
H^2(\mathfrak{c};\mathfrak{c})=\frac{\text{ker }d:C^2(\mathfrak{c};\mathfrak{c})\longrightarrow C^3(\mathfrak{c};\mathfrak{c})}{\text{im }d:C^1(\mathfrak{c};\mathfrak{c})\longrightarrow C^2(\mathfrak{c};\mathfrak{c})}=\frac{C^2(\mathfrak{c};\mathfrak{c})}{C^2(\mathfrak{c};\mathfrak{c})}=0.
\end{equation}
This means that the CKY superalgebras of constant curvature manifolds are rigid Lie superalgebras and their bilinear brackets cannot be deformed to obtain other Lie superalgebras. Hence, the CKY superalgebras are geometric invariants for constant curvature backgrounds and they are more general invariants than KY superalgebras since they are some kind of extensions of KY superalgebras to more general objects.

\section{Example: $AdS$ Spacetimes}

Because of their importance in string theory and supergravity, we will consider the $AdS$ spacetimes as examples of constant curvature manifolds. We can define the KY and CKY superalgebras from KY and CKY forms of maximal numbers on those backgrounds. An $n$-dimensional $AdS$ spacetime $AdS_n$ can be realized as a hyperboloid in an $(n+1)$-dimensional flat Minkowski spacetime. Moreover, KY forms and CKY forms of Minkowski and $AdS$ spacetimes are related to each other.

For an $(n+1)$-dimensional Minkowski spacetime $M^{n+1}$, the number of KY $p$-forms is maximal and is given by
\begin{equation}
K_p=\left(
             \begin{array}{c}
               n+2 \\
               p+1 \\
             \end{array}
           \right)=\left(
             \begin{array}{c}
               n+1 \\
               p \\
             \end{array}
           \right)+\left(
             \begin{array}{c}
               n+1 \\
               p+1 \\
             \end{array}
           \right).
\end{equation}
The first term of the sum on the right hand side that is $\left(
             \begin{array}{c}
               n+1 \\
               p \\
             \end{array}
           \right)$ corresponds to the number of translational KY $p$-forms of the Minkowski spacetime. The number of boost/rotational KY $p$-forms is the second term in the sum that is $\left(
             \begin{array}{c}
               n+1 \\
               p+1 \\
             \end{array}
           \right)$. Because of the hyperboloid structure of $AdS_n$ in $M^{n+1}$, the boost/rotational KY $p$-forms of $M^{n+1}$ are tangent to the hyperboloid $AdS_n$ and they correspond to all of the KY $p$-forms of $AdS_n$. Moreover, the translational KY $p$-forms of $M^{n+1}$ are equivalent to the CKY $p$-forms of $AdS_n$ which are not KY forms. Indeed, this means that the number of KY $p$-forms in $M^{n+1}$ is equivalent to the number of CKY $p$-forms in $AdS_n$. So, the number of KY $p$-forms of $AdS_n$ is the number of boost/rotational KY $p$-forms of $M^{n+1}$ and the number of CKY $p$-forms of $AdS_n$ is the number of all KY $p$-forms of $M^{n+1}$. The KY superalgebra $\mathfrak{k}$ is a subsuperalgebra of the CKY superalgebra $\mathfrak{c}$ and we can write
\[
\mathfrak{c}=\mathfrak{k}\oplus\mathfrak{m}
\]
where $\mathfrak{m}$ is the subset of CKY forms that do not correspond to KY forms. The relation between the total dimensions of these algebras in $n$-dimensions is exactly given by (36).

We will analyze the dimension and structure of KY and CKY superalgebras in lower dimensional $AdS$ spacetimes. They consist of geometric invariants of the backgrounds because of the rigidity property of them.

i) $AdS_3$. The metric of $AdS_3$ spacetime is written as
\[
ds^2=-\left(\frac{r^2}{l^2}+1\right)dt^2+\left(\frac{r^2}{l^2}+1\right)^{-1}dr^2+r^2d\phi^2
\]
in terms of the coordinates $t,r,\phi$ and the cosmological length $l$ related to the cosmological constant $\Lambda=-1/l^2$. It can be written in a locally Lorentzian form
\[
ds^2=-e^0\otimes e^0+e^1\otimes e^1+e^2\otimes e^2
\]
by defining the coframe basis
\[
e^0=\left(\frac{r^2}{l^2}+1\right)^{1/2}dt\quad,\quad e^1=\left(\frac{r^2}{l^2}+1\right)^{-1/2}dr\quad,\quad e^2=rd\phi.
\]
In three dimensions, the number of different degree KY forms are given as

\quad\\
{\centering{
\begin{tabular}{c|c c}

$p$ & \quad$1$\quad & \quad$2$ \\ \hline
$K_p$ & \quad$6$\quad & \quad$4$ \\

\end{tabular}}
\quad\\
\quad\\
\quad\\}

So, the dimension of the KY superalgebra is $(K_{\text{odd}}\big|K_{\text{even}})=(6\big|4)$. The constant functions and the volume form with constant coefficients are trivial solutions of the KY equation and we do not consider them in the algebra since they trivially commute with all other elements. The explicit forms of KY $p$-forms of $AdS_3$ are obtained in \cite{Ertem Acik}. By choosing a convenient basis, KY 1-forms of $AdS_3$ can be written as
\begin{eqnarray}
\alpha_1&=&-\left(\frac{r^2}{l^2}+1\right)^{1/2}e^0+re^2\nonumber\\
\alpha_2&=&-\frac{r}{l^2}\left(\cosh{\frac{t}{l}}\cos{\phi}+\sinh{\frac{t}{l}}\sin{\phi}\right)e^0-\frac{1}{l}\left(\sinh{\frac{t}{l}}\cos{\phi}+\cosh{\frac{t}{l}}\sin{\phi}\right)e^1\nonumber\\
&&+\frac{1}{l}\left(\frac{r^2}{l^2}+1\right)^{1/2}\left(\sinh{\frac{t}{l}}\sin{\phi}-\cosh{\frac{t}{l}}\cos{\phi}\right)e^2\nonumber\\
\alpha_3&=&-\frac{r}{l^2}\left(\sinh{\frac{t}{l}}\cos{\phi}+\cosh{\frac{t}{l}}\sin{\phi}\right)e^0-\frac{1}{l}\left(\cosh{\frac{t}{l}}\cos{\phi}+\sinh{\frac{t}{l}}\sin{\phi}\right)e^1\nonumber\\
&&+\frac{1}{l}\left(\frac{r^2}{l^2}+1\right)^{1/2}\left(\cosh{\frac{t}{l}}\sin{\phi}+\sinh{\frac{t}{l}}\cos{\phi}\right)e^2\nonumber\\
\beta_1&=&\left(\frac{r^2}{l^2}+1\right)^{1/2}e^0+re^2\nonumber\\
\beta_2&=&-\frac{r}{l^2}\left(\cosh{\frac{t}{l}}\cos{\phi}-\sinh{\frac{t}{l}}\sin{\phi}\right)e^0-\frac{1}{l}\left(\sin{\frac{t}{l}}\cos{\phi}-\cosh{\frac{t}{l}}\sin{\phi}\right)e^1\nonumber\\
&&+\frac{1}{l}\left(\frac{r^2}{l^2}+1\right)^{1/2}\left(\sinh{\frac{t}{l}}\sin{\phi}+\cos{\frac{t}{l}}\cos{\phi}\right)e^2\nonumber\\
\beta_3&=&-\frac{r}{l^2}\left(\sinh{\frac{t}{l}}\cos{\phi}-\cosh{\frac{t}{l}}\sin{\phi}\right)e^0-\frac{1}{l}\left(\cosh{\frac{t}{l}}\cos{\phi}-\sinh{\frac{t}{l}}\sin{\phi}\right)e^1\nonumber\\
&&+\frac{1}{l}\left(\frac{r^2}{l^2}+1\right)^{1/2}\left(\cosh{\frac{t}{l}}\sin{\phi}-\sinh{\frac{t}{l}}\cos{\phi}\right)e^2.\nonumber
\end{eqnarray}
These 1-forms correspond to the metric duals of Killing vector fields and they satisfy a $\mathfrak{so}(2,2)$ Lie algebra structure under the SN bracket. Moreover, $\alpha_i$ and $\beta_j$ commute with each other and we have the isometry algebra of $AdS_3$ as
\[
\mathfrak{so}(2,2)=\mathfrak{sl}(2,\mathbb{R})\oplus\mathfrak{sl}(2,\mathbb{R}).
\]
KY 2-forms of $AdS_3$ are
\begin{eqnarray}
\omega_1&=&\cos{\phi}e^0\wedge e^1-\left(\frac{r^2}{l^2}+1\right)^{1/2}\sin{\phi}e^0\wedge e^2\nonumber\\
\omega_2&=&\sin{\phi}e^0\wedge e^1+\left(\frac{r^2}{l^2}+1\right)^{1/2}\cos{\phi}e^0\wedge e^2\nonumber\\
\omega_3&=&-\frac{1}{l}\sinh{\frac{t}{l}}e^0\wedge e^2+\cosh{\frac{t}{l}}e^1\wedge e^2\nonumber\\
\omega_4&=&-\frac{1}{l}\cosh{\frac{t}{l}}e^0\wedge e^2+\sinh{\frac{t}{l}}e^1\wedge e^2.\nonumber
\end{eqnarray}
As can be seen from direct computation, all KY 2-forms commute with each other under the SN bracket
\[
[\omega_i,\omega_j]_{SN}=0
\]
for all $i,j,=1,2,3,4$. The SN brackets between KY 1-forms and KY 2-forms are given by
\begin{eqnarray}
[\alpha_1,\omega_2]_{SN}=[\alpha_2,\omega_4]_{SN}=[\alpha_3,\omega_3]_{SN}&=&\,\,\,\,\omega_1=[\beta_1,\omega_2]_{SN}=[\beta_2,\omega_4]_{SN}=[\beta_3,\omega_3]_{SN}\nonumber\\
{[\alpha_1,\omega_1]}_{SN}=[\alpha_2,\omega_3]_{SN}=[\alpha_3,\omega_4]_{SN}&=&-\omega_2={[\beta_1,\omega_1]}_{SN}=[\beta_2,\omega_3]_{SN}=[\beta_3,\omega_4]_{SN}\nonumber\\
{[\alpha_1,\omega_4]}_{SN}=[\alpha_2,\omega_2]_{SN}=[\alpha_3,\omega_1]_{SN}&=&\,\,\,\,\omega_3={[\beta_1,\omega_4]}_{SN}=[\beta_2,\omega_2]_{SN}=[\beta_3,\omega_1]_{SN}\nonumber\\
{[\alpha_1,\omega_3]}_{SN}=[\alpha_2,\omega_1]_{SN}=[\alpha_3,\omega_2]_{SN}&=&-\omega_4={[\beta_1,\omega_3]}_{SN}=[\beta_2,\omega_1]_{SN}=[\beta_3,\omega_2]_{SN}.\nonumber
\end{eqnarray}
Then, we obtain the structure of the KY superalgebra as $\mathfrak{k}=\mathfrak{k}_{\bar{0}}\oplus\mathfrak{k}_{\bar{1}}=\mathfrak{so}(2,2)\oplus\mathbb{R}^4$ in $AdS_3$. The even part $\mathfrak{k}_{\bar{0}}$ corresponds to the isometry algebra $\mathfrak{so}(2,2)$, the odd part $\mathfrak{k}_{\bar{1}}$ consists of the vector space of KY 2-forms and the action of $\mathfrak{k}_{\bar{0}}$ on $\mathfrak{k}_{\bar{1}}$ is given by the above brackets. Since it cannot be deformed to another Lie superalgebra, it corresponds to a geometric invariant of $AdS_3$.

On the other hand, the number of CKY $p$-forms in $AdS_3$ are given by

\quad\\
{\centering{
\begin{tabular}{c|c c}

$p$ & \quad$1$\quad & \quad$2$ \\ \hline
$C_p$ & \quad$10$\quad & \quad$10$ \\

\end{tabular}}
\quad\\
\quad\\
\quad\\}

and calculating the explicit forms of CKY forms is more difficult than the KY forms case. The dimension of the CKY superalgebra is $(C_{\text{odd}}\big|C_{\text{even}})=(10\big|10)$.

ii) $AdS_4$. The explicit calculation of KY $p$-forms in a class of four dimensional spherically symmetric spacetimes including $AdS_4$ has been done in \cite{Acik Ertem Onder Vercin3}. The number of KY $p$-forms of $AdS_4$ are

\quad\\
{\centering{
\begin{tabular}{c|c c c}

$p$ & \quad$1$\quad & \quad$2$\quad & \quad$3$ \\ \hline
$K_p$ & \quad$10$\quad & \quad$10$\quad & \quad$5$ \\

\end{tabular}}
\quad\\
\quad\\
\quad\\}

and the dimension of the KY superalgebra is $(K_{\text{odd}}\big|K_{\text{even}})=(15\big|10)$. It is an extension of the isometry algebra $\mathfrak{so}(3,2)$. For CKY $p$-forms, we have

\quad\\
{\centering{
\begin{tabular}{c|c c c}

$p$ & \quad$1$\quad & \quad$2$\quad & \quad$3$ \\ \hline
$C_p$ & \quad$15$\quad & \quad$20$\quad & \quad$15$ \\

\end{tabular}}
\quad\\
\quad\\
\quad\\}

and the dimension of the CKY superalgebra is $(C_{\text{odd}}\big|C_{\text{even}})=(30\big|20)$.

iii) $AdS_5$. In five dimensions, we have the number of KY $p$-forms in $AdS_5$ as

\quad\\
{\centering{
\begin{tabular}{c|c c c c}

$p$ & \quad$1$\quad & \quad$2$\quad & \quad$3$\quad & \quad$4$ \\ \hline
$K_p$ & \quad$15$\quad & \quad$20$\quad & \quad$15$\quad & \quad$6$ \\

\end{tabular}}
\quad\\
\quad\\
\quad\\}
with the dimension of the KY superalgebra $(K_{\text{odd}}\big|K_{\text{even}})=(30\big|26)$. The number of CKY $p$-forms are

\quad\\
{\centering{
\begin{tabular}{c|c c c c}

$p$ & \quad$1$\quad & \quad$2$\quad & \quad$3$\quad & \quad$4$ \\ \hline
$C_p$ & \quad$21$\quad & \quad$35$\quad & \quad$35$\quad & \quad$21$ \\

\end{tabular}}
\quad\\
\quad\\
\quad\\}

and the dimension of the CKY superalgebra is $(C_{\text{odd}}\big|C_{\text{even}})=(56\big|56)$. The dimensional analysis for higher dimensions can be done similarly. The isometry algebra of $AdS_n$ is $\mathfrak{so}(n-1,2)$ and the KY superalgebra extends the structure of it, while CKY superalgebras correspond to more general invariants that include them.

\section{Conclusion}

The symmetry algebras of Killing vector fields and conformal Killing vector fields have generalizations to hidden symmetry superalgebras in constant curvature manifolds. While the bilinear operation of KY superalgebras can be defined from the generalization of the Lie bracket of vector fields to the SN bracket, a new CKY bracket, which reduces to the SN bracket in special cases, has to be defined for the case of CKY superalgebras. By analyzing the Lie superalgebra cohomology theories of KY and CKY superalgebras, we have shown that the second cohomology groups of them are trivial and they cannot be deformed to obtain other Lie superalgebras. This means that, in constant curvature manifolds, KY and CKY superalgebras correspond to new geometric invariants and can be used to characterize the background manifolds. Since CKY superalgebras can also be defined in Einstein manifolds by using the normal CKY forms, they can also be used in the characterization of them. Although finding the explicit forms of KY and CKY forms in various backgrounds is not an easy task, the dimension of the superalgebra in constant curvature case can be calculated. We give an explicit example of the KY superalgeba in $AdS_3$ and find the dimensions of the superalgebras in $AdS_4$ and $AdS_5$. They can be generalized to higher dimensions.

Killing superalgebras of supergravity backgrounds are constructed out of Killing vector fields and Killing spinors by using the Lie bracket of vector fields, Lie derivative on spinors and the Dirac currents of spinors. Although the extensions of them to include KY forms can be defined by using the SN bracket, generalized symmetry operators of Killing spinors and $p$-form Dirac currents of Killing spinors in constant curvature spacetimes \cite{Ertem1}, obtaining a Lie superalgebra structure from these extensions is still an open problem. By investigating the relations between KY superagebras and extended Killing superalgebras and their cohomology theories, one can obtain new developments in this field. Similar discussions can be done for CKY superalgebras and extended conformal superalgebras which are constructed out of CKY forms and twistor spinors by using the CKY bracket, generalized symmetry operators of twistor spinors and $p$-form Dirac currents of twistor spinors in constant curvature and Einstein manifolds \cite{Ertem2}. These investigations may be extended to obtain geometric invariants of more general manifolds.


\end{document}